\documentclass[draftcls,onecolumn]{IEEEtran}

\usepackage{graphicx}
\usepackage[utf8]{inputenc} 
\usepackage[T1]{fontenc}
\usepackage{url}
\usepackage{ifthen}
\usepackage{cite}
\usepackage{color}
\usepackage{amsthm}

\usepackage{cite}
\usepackage{graphicx}
\usepackage{hyperref}
\usepackage{amsmath}
\usepackage{amssymb}
\usepackage{bm}
\usepackage{bbm}
\usepackage{algorithmicx}
\usepackage{algorithm}
\usepackage{algpseudocode}
\usepackage{array}
\usepackage{booktabs}
\usepackage{multirow}
\usepackage{makecell}
\usepackage{mathrsfs}
\newtheorem{conjecture}{Conjecture}
\newtheorem{definition}{Definition}
\newtheorem{lemma}{Lemma}
\newtheorem{theorem}{Theorem}
\newtheorem{corollary}{Corollary}
\newtheorem{proposition}{Proposition}

\newcommand \R{\bm{R}}

\newcommand \setS{\mathcal{S}}

\newcommand \setC{\mathcal{C}}
\newcommand \Z{\mathbb{Z}^+}
\newcommand \E{\mathbb{E}}
\newcommand \Prob{\mathscr{P}}
\newcommand \lalpha{\overline{\alpha}}

\newcommand {\I}{\mathbbm{1}}

\begin{document}
\title{On the Most Informative Boolean Functions\\ of the Very Noisy Channel} 


\author{\IEEEauthorblockN{Hengjie~Yang and Richard~D.~Wesel}
\IEEEauthorblockA{
\\Department of Electrical and Computer Engineering\\
University of California, Los Angeles, Los Angeles, CA 90095, USA\\
Email: \{hengjie.yang, wesel\}@ucla.edu}
}


\maketitle

\begin{abstract}
Let $X^n$ be a uniformly distributed $n$-dimensional binary vector, and $Y^n$ be the result of passing $X^n$ through a binary symmetric channel (BSC) with crossover probability $\alpha$. A recent conjecture postulated by Courtade and Kumar states that for any Boolean function $f:\{0,1\}^n\to\{0,1\}$, $I(f(X^n);Y^n)\le 1-H(\alpha)$. Although the conjecture has been proved to be true in the dimension-free high noise regime by Samorodnitsky, here we present a calculus-based approach to show a dimension-dependent result by examining the second derivative of $H(\alpha)-H(f(X^n)|Y^n)$ at $\alpha=1/2$. Along the way, we show that the dictator function is the most informative function in the high noise regime.
\end{abstract}

\section{Introduction}
\subsection{Previous Work}
In \cite{Courtade_2014}, Courtade and Kumar postulated the following most informative Boolean function conjecture:
\begin{conjecture}
\label{conjecture: mutual info}
Let $X^n=(X_1,\dots,X_n)$ be a sequence of $n$ i.i.d. Bernoulli $(1/2)$ random variables, and let $Y^n$ be the result of passing $X^n$ through a memoryless binary symmetric channel (BSC) with crossover probability $\alpha$. For any Boolean function $f:\{0,1\}^n\to\{0,1\}$, we have
\begin{align}
    I(f(X^n);Y^n)\le 1-H(\alpha).\label{eq:conjecture 1}
\end{align}
\end{conjecture}

Intuitively, Conjecture \ref{conjecture: mutual info} asks the following question: ``\emph{What is the most informative bit $X^n$ can provide about $Y^n$?}'' It can be readily verified that the dictator function, $f(X^n)=X_i, i\in\{1,2,\cdots,n\}$, can achieve the equality in \eqref{eq:conjecture 1}, suggesting that the dictator function might be the most informative function that $X^n$ can reveal about $Y^n$. However, the rigorous proof of showing that the dictator function is indeed the most informative function is still elusive. Courtade and Kumar \cite{Courtade_2014} showed that Conjecture \ref{conjecture: mutual info} holds when $\alpha\to0$ using an edge-isoperimetric argument.

The first result pertaining to Conjecture \ref{conjecture: mutual info} dates back to the work of Wyner and Ziv \cite{Wyner_1973}, known as Mrs. Gerber's Lemma.

\begin{theorem}[Mrs. Gerber's Lemma]
Let $X^n, Y^n$ be binary random-$n$ vectors, which are input and output, respectively, of a binary symmetric channel with crossover probability $\alpha$. Let $h(x)\triangleq -x\log x-(1-x)\log(1-x), x\in[0,1]$ be the binary entropy function. Let $H(X^n), H(Y^n)$ be the entropies of $X^n, Y^n$, respectively, with $H(X^n)$ satisfying $\frac1nH(X^n)\ge h(\pi_0), 0\le \pi_0\le 1$. Then
\begin{align}
\frac1nH(Y^n)\ge h(\pi_0(1-\alpha)+(1-\pi_0)\alpha),
\end{align}
with equality if and only if $X^n=(X_1,X_2,\cdots,X_n)$ are independent and with $H(X_i)=h(\pi_0), i\in\{1,2,\cdots,n\}.$
\end{theorem}

With Mrs. Gerber's Lemma, Erkip \cite{Erkip_1998} showed the following universal upper bound on $I(f(X^n);Y^n)$
\begin{align}
I(f(X^n);Y^n)\le(1-2\alpha)^2,\quad \forall \alpha\in[0,1]. \label{eq:Erkip}
\end{align}
 However, \eqref{eq:Erkip} is still strictly weaker than \eqref{eq:conjecture 1}.

In \cite{Ordentlich_2016}, Ordentlich, Shayevitz, and Weinstein used Fourier analytic techniques and leveraged hypercontractivity to improve the upper bound on $I(f(X^n);Y^n)$ for all balanced Boolean functions, i.e., Boolean functions satisfying $\Prob\{f(X^n)=0\}=\Prob\{f(X^n)=1\}=\frac12$, which beats Erkip's bound in \eqref{eq:Erkip} when $\alpha>\frac13$. 
\begin{theorem}[Ordentlich \emph{et al.}]
For any balanced Boolean function $f:\{0,1\}^n\to\{0,1\}$, and any $\frac12(1-\frac1{\sqrt{3}})\le \alpha\le \frac12$, we have that
\begin{align}
I(f(X^n);Y^n)\le\frac{\log_2e}{2}(1-2\alpha)^2+9(1-\frac{\log_2e}{2})(1-2\alpha)^4.
\end{align}
\end{theorem}
As a corollary, they also showed that the dictator function is the most informative balanced function in the high noise regime.

So far, the most promising result is due to Samorodnitsky \cite{Samorodnitsky_2016} who proved that Conjecture \ref{conjecture: mutual info} holds in the dimension-free high noise regime, i.e., for $\alpha\in(\frac12-\delta, \frac12+\delta)$ with $\delta>0$ being a dimension independent number, by considering the entropy of the image of $f$ under a noise operator.
\begin{theorem}[Samorodnitsky]
There exists an absolute $\delta>0$ such that for any noise $\alpha>0$ with $(1-2\alpha)^2<\delta$ and for any Boolean function $f:\{0,1\}^n\to\{0,1\}$, we have
\begin{align}
I(f(X^n);Y^n)\le 1-H(\alpha).
\end{align}
\end{theorem}

In addition to Conjecture \ref{conjecture: mutual info}, some related conjectures are also addressed in \cite{Courtade_2014}. One conjecture is that, for Boolean functions $f_1, f_2$, does it hold that 
\begin{align}
I(f_1(X^n);f_2(Y^n))\le 1-H(\alpha)? \label{eq: double booleans}
\end{align}
This conjecture is then positively resolved by Pichler, Matz, and Piantanida \cite{Pichler_2016} using Fourier-analytic arguments. The Gaussian analogy of Conjecture \ref{conjecture: mutual info} is proved by Kindler, O'Donnell, and Witmer \cite{Kindler_2015}. Anantharam \emph{et al.} \cite{Anatharam_2013} conjectured a result related to the chordal slope of the hypercontractivity ribbon of a pair of binary random variables, which would imply \eqref{eq: double booleans}. However, this stronger result still remains open.

Recently, a complementary problem concerning Conjecture \ref{conjecture: mutual info} is posed and proved by Huleihel and Ordentlich \cite{Huleihel_2017}.
\begin{theorem}[Huleihel \emph{et al.}]
For any function $f:\{0,1\}^n\to\{0,1\}^{n-1}$, we have
\begin{align}
I(f(X^n);Y^n)\le (n-1)(1-H(\alpha)),
\end{align}
and this bound is attained with equality by, e.g., $f(x^n)=(x_1,x_2,\cdots,x_{n-1}).$
\end{theorem}

Li and M\'edard \cite{Li_2018} studied the problem of maximizing the $p$-th moment of the image of $f$ under noise operator and discussed the connection between noise stability and Conjecture \ref{conjecture: mutual info}.

\subsection{Main Contributions}
In this paper, we prove that Conjecture \ref{conjecture: mutual info} holds in the high noise regime by  applying a calculus-based approach directly to \eqref{eq:conjecture 1}. Namely, by defining $F_f(\alpha)\triangleq H(\alpha)-H(f(X^n)|Y^n)$ and $T(\Prob\{f(X^n)=0\})\triangleq 1-H(f(X^n))$, we prove that given dimension $n$, for any Boolean function $f:\{0,1\}^n\to\{0,1\}$, $F_f(\alpha)\le T(\Prob\{f(X^n)=0\})$ holds for $\alpha\in(\frac12-\delta_n, \frac12+\delta_n)$, where $\delta_n$ is some positive constant dimensionally dependent on $n$. Algebraically, we will show that $F_f(\alpha)$ is always a concave function in the high noise regime by demonstrating that $F_f''(1/2)$ is always nonpositive.

For a given $n$ and $M$, let us consider the set of Boolean functions $\{f: |f^{-1}(0)|=M\}$. We say ``Boolean function $f$ is most informative in the high noise regime'' if $I(f(X^n);Y^n)$ is undominated in small interval centered at $1/2$, i.e. if $F_f(\alpha)$ for this choice of $f$ is greater than or equal to $F_f(\alpha)$ for any other choice of $f$ in the set. Since $F_f(1/2)=T, F_f'(1/2)=0$ for any Boolean function $f$, the most informative condition reduces to finding $f$ that maximizes $F_f''(1/2)$.

As pointed out in \cite{Courtade_2014}, Conjecture \ref{conjecture: mutual info} inherently consists of two components, the \emph{structure} of maximally-informative Boolean functions that asks which Boolean function maximizes $I(f(X^n);Y^n)$ for a given $n$ and fixed $|f^{-1}(0)|$, and the \emph{inequality component} of the conjecture that concerns how to establish the inequality given that $f$ is the maximally-informative function for a given $n$ and fixed $|f^{-1}(0)|$. Indeed, the maximization of $F_f''(1/2)$ requires the solution to the above two components. In this paper, we introduce the notion of \emph{ratio spectrum} of $f^{-1}(0)$,  an integer sequence that characterizes the structure of $f^{-1}(0)$, which uniquely determines $F_f''(1/2)$. We then show that the lex function (all notions will be defined shortly) is \emph{a} maximizing Boolean function for a given $n$ and fixed $|f^{-1}(0)|<2^{n-1}$ and the dictator function (a special case $f(X^n)=X_1$ of which is lex) is \emph{the} maximizing Boolean function for $|f^{-1}(0)|=2^{n-1}$. Finally, we establish the nonpositivity of $F''_f(1/2)$ given that $f$ is lex and show that the dictator function is the only type of functions that achieve $F_f''(1/2)=0$, thus implying its optimality.


This paper is organized as follows. Sec. \ref{sec:reformulation and main results} introduces the reformulation of Conjecture \ref{conjecture: mutual info} and gives our main results. Sec. \ref{sec: proof of lemma 1} and Sec. \ref{sec: proof of lemma 2} present the proofs of our main results. Sec. \ref{sec: discussion} discusses the limitations and open problems of our calculus-based approach.

\section{Reformulation and Main Results}\label{sec:reformulation and main results}
\subsection{Reformulation}
Let $\setS$ denote the universal set of all $n$-dimensional binary sequences with $S=|\setS|=2^n$. The logarithm base is $2$, whereas in Sec. \ref{sec: proof of lemma 2}, we assume natural logarithm. For a scalar $p\in[0,1]$, $H(p)\triangleq -p\log p-(1-p)\log(1-p)$ denotes the binary entropy function.


\begin{figure}[t]
\centering
\includegraphics[scale=0.8]{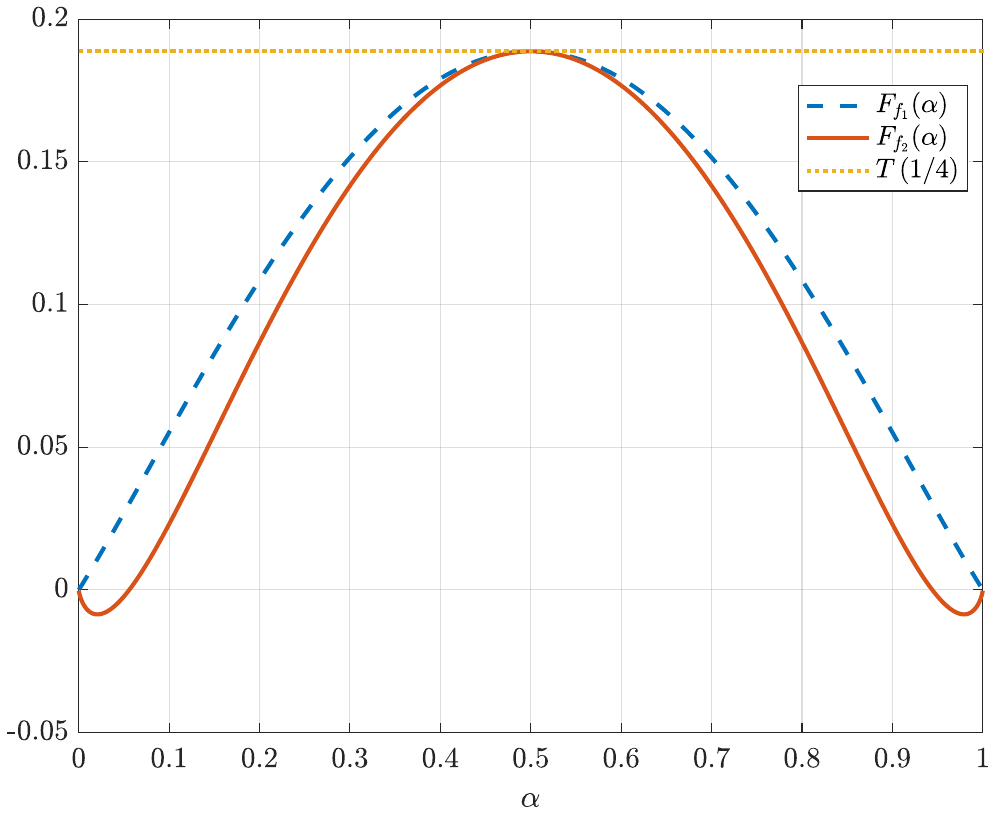}
\caption{An example of $F_f(\alpha)$ and $T(\mathscr{P}\{f(X^n)=0\})$, with $n=4, |f^{-1}(0)|=4, \mathscr{P}\{f(X^n)=0\}=1/4$. We show two typical shapes of $F_f(\alpha)$: the quasi-concave shape for $F_{f_1}(\alpha)$ with $f^{-1}_1(0)=\{0,1,2,3\}$ and the ``single-peak wave'' shape for $F_{f_2}(\alpha)$ with $f^{-1}_2(0)=\{0,1,2,4\}$, where the element in the curly bracket represents the decimal representation of an $n$-dimensional binary sequence.}
\label{fig:example of F and T}
\end{figure}

Define 
\begin{align}
F_f(\alpha)\triangleq&H(\alpha)-H(f(X^n)|Y^n)\notag\\
  =&H(\alpha)-\E_{Y^n} H(\Prob\{f(X^n)=0|Y^n\})\notag\\
  =&H(\alpha)-\frac1S\sum_{y^n\in\setS}H(\Prob\{f(X^n)=0|y^n,\alpha\}),\label{eq: definition of F_f}
\end{align}
and
\begin{align}
T(\mathscr{P}\{f(X^n)=0\})\triangleq1-H(f(X^n))=1-H\left(\frac{|f^{-1}(0)|}{2^n}\right),
\end{align}
where by letting $\lalpha=1-\alpha$,
\begin{align}
\Prob\{f(X^n)=0|y^n,\alpha\}=&\sum_{x^n\in f^{-1}(0)}\alpha^{d(x^n,y^n)}\lalpha^{n-d(x^n,y^n)}.
\end{align}

With the above notation, Conjecture \ref{conjecture: mutual info} translates to the following conjecture.
\begin{conjecture}
Given dimension $n$, for any Boolean function $f:\{0,1\}^n\to\{0,1\}$, we have
\begin{align}
\max_{\alpha\in[0,1]}F_f(\alpha)= F_f\left(\frac12\right)=T(\mathscr{P}\{f(X^n)=0\}).
\end{align}
\end{conjecture}

Note that it is trivial to show that $F_f\left(1/2\right)=T(\mathscr{P}\{f(X^n)=0\})$ since when $\alpha=1/2$, $X^n$ and $Y^n$ are independent, which implies that $F_f\left(1/2\right)=H(1/2)-H(f(X^n))=T(\mathscr{P}\{f(X^n)=0\})$. Therefore, establishing that $\max_{\alpha\in[0,1]}F_f(\alpha)= F_f(1/2)$ for any Boolean function $f$ is the key to solving Conjecture \ref{conjecture: mutual info}. Also worth mentioning is the dictator function, $f(X^n)=X_i, i\in\{1,2,\cdots,n\}$, for which $\mathscr{P}\{f(X^n)=0\}=1/2$, $F_f(\alpha)=0$ for $\alpha\in[0,1]$ and $T(\mathscr{P}\{f(X^n)=0\})=0$.

As an example of the reformulation, Fig. \ref{fig:example of F and T} shows that $F_f(\alpha)\le T(\mathscr{P}\{f(X^n)=0\})$ for $n=4$ and $|f^{-1}(0)|=4$. Meanwhile, Fig. \ref{fig:example of F and T}. also depicts two typical shapes of $F_f(\alpha)$: a quasi-concave shape as shown by $F_{f_1}(\alpha)$, and  a ``single-peak wave'' shape as shown by $F_{f_2}(\alpha)$. In fact, we conjecture that these are the only two possible shapes of $F_f(\alpha), \alpha\in(0,1)$. Note that even for dictator function $f(X^n)=X_i, i\in\{1,2,\cdots,n\}$, $F_f(\alpha)=0$ is still quasi-concave.

\subsection{Main Results}
Our main result is the following theorem.
\begin{theorem}\label{theorem: main result}
For a given $n$, there exists a constant $\delta_n>0$ such that, for any Boolean function $f:\{0,1\}^n\to\{0,1\}$, 
\begin{align}
\max_{\alpha\in(\frac12-\delta_n,\frac12+\delta_n)}F_f(\alpha)=F_f\left(\frac12\right)=T(\mathscr{P}\{f(X^n)=0\}).
\end{align}
\end{theorem}

The entire paper is to establish Theorem \ref{theorem: main result} by proving the following lemmas.
\begin{lemma}\label{lemma: symmetry}
Given $n$, for any Boolean function $f:\{0,1\}^n\to\{0,1\}$, $F_f(\alpha)$ is symmetric with respect to $\alpha=1/2$.
\end{lemma}

As a corollary, combined with the fact that $F_f(\alpha)$ is differentiable, Lemma \ref{lemma: symmetry} also implies that 
\begin{align}
F_f'\left(\frac12\right)=0.
\end{align} 
\begin{lemma}\label{lemma: second_deriv}
Given dimension $n$, for any Boolean function $f:\{0,1\}^n\to\{0,1\}$, we have
\begin{align}
F_f''\left(\frac12\right)\le0,
\end{align}
with equality if and only if $f$ is the dictator function.
\end{lemma}

To see how Lemma \ref{lemma: symmetry} and Lemma \ref{lemma: second_deriv} lead to Theorem \ref{theorem: main result}, notice that for each Boolean function 
$f:\{0,1\}^n\to\{0,1\}$, Lemma \ref{lemma: symmetry} and Lemma \ref{lemma: second_deriv} imply that there exists a $\delta_{n,f}>0$ such that $F_f(\alpha)\le F_f(1/2)=T(\Prob\{f(X^n)=0\})$ holds for $\alpha\in(1/2-\delta_{n,f},1/2+\delta_{n,f})$. Since the number of all possible Boolean functions is finite for a fixed $n$, choosing $\delta_n=\min_{f:|f^{-1}(0)|\le2^n}\{\delta_{n,f}\}$ immediately implies Theorem \ref{theorem: main result}.

Lemma \ref{lemma: second_deriv} suggests that for any Boolean function $f$, $F_f(\alpha)$ always resembles a concave function in the high noise regime. Also, by our previous most informative Boolean function argument, Lemma \ref{lemma: second_deriv} clearly demonstrates that the dictator function is the most informative Boolean function in the high noise regime.


\section{Proof of Lemma 1}\label{sec: proof of lemma 1}
Consider $\alpha\in[0,1/2]$ and its symmetric part $\lalpha=1-\alpha\in[1/2,1]$. We have
\begin{align}
\Prob\{f(X^n)=0|y^n,\alpha\}=&\sum_{x^n\in f^{-1}(0)}\alpha^{d(x^n,y^n)}\lalpha^{n-d(x^n,y^n)}\notag\\
  =&\sum_{x^n\in f^{-1}(0)}\alpha^{n-d(x^n,\bar{y}^n)}\lalpha^{d(x^n,\bar{y}^n)}\notag\\
  =&\Prob\{f(X^n)=0|\bar{y}^n,\lalpha\}
\end{align}
where $\bar{y_i}=1-y_i, i=\{1,2,\cdots,n\}$. Hence,
\begin{align}
F_f(\alpha)=&H(\alpha)-\frac1S\sum_{y^n\in\setS}H(\Prob\{f(X^n)=0|y^n,\alpha\})\notag\\
  =&H(\lalpha)-\frac1S\sum_{y^n\in\setS}H(\Prob\{f(X^n)=0|\bar{y}^n,\lalpha\})\notag\\
  =&F_f(\lalpha)
\end{align}
which completes the proof of Lemma \ref{lemma: symmetry}.

Similarly, an additional symmetry property for \emph{complementary function} $1-f$ with respect to $f$ is presented as follows.

\begin{theorem}\label{theorem:complementary function}
For any Boolean function $f$, define its complementary function $f^c=1-f$. We have $F_f(\alpha)=F_{1-f}(\alpha)$.
\end{theorem}
\begin{IEEEproof}
By $H(p)=H(1-p), p\in[0,1]$,
\begin{align}
F_f(\alpha)=&H(\alpha)-\frac1S\sum_{y^n\in\setS}H(1-\Prob\{f(X^n)=1|y^n,\alpha\})\notag\\
  =&H(\alpha)-\frac1S\sum_{y^n\in\setS}H(\Prob\{f^c(X^n)=1|y^n,\alpha\})\notag\\
  =&F_{1-f}(\alpha)
\end{align}
\end{IEEEproof}

The implication of Theorem \ref{theorem:complementary function} is that it suffices to focus on $1\le |f^{-1}(0)|\le 2^{n-1}$ for a given $n$.

\section{Proof of Lemma 2}\label{sec: proof of lemma 2}

The proof of Lemma \ref{lemma: second_deriv} proceeds as follows: first, for a given $n$ and $M$, $1\le M\le 2^{n-1}$, for Boolean function $f$ with $|f^{-1}(0)|=M$, we derive the general formula for $F_f''(1/2)$, which is uniquely determined by the \emph{ratio spectrum} of $f^{-1}(0)$. Next, we prove that $\max_{f: |f^{-1}(0)|=M}F_f''(1/2)\le0$ for any $1\le M\le 2^{n-1}$. Note that the entire proof now assumes natural logarithms unless otherwise specified. 

During the proof, one can see that the \emph{lex function} is \emph{a} locally most informative function among all $f$ with $|f^{-1}(0)|=M$, achieving $\max_{f:|f^{-1}(0)|=M}F_f''(1/2)$. In particular, if $M<2^{n-1}$, we have $\max_{f:|f^{-1}(0)|=M}F_f''(1/2)<0$. If $M=2^{n-1}$, lex function becomes dictator function $f(X^n)=X_1$, achieving $\max_{f:|f^{-1}(0)|=2^{n-1}}F_f''(1/2)=0$. By the uniqueness of the dictator function, we conclude that the dictator function is the most informative function in the high noise regime.

We first introduce several new definitions which will play an important role in proving Lemma \ref{lemma: second_deriv}. For brevity, let $M\triangleq |f^{-1}(0)|$ henceforth unless otherwise specified.
\begin{definition}[lex function]
Boolean function $f$ is said to be lex if $f^{-1}(0)$ only contains the first $M$ lexicographically ordered $n$-dimensional binary sequences.
\end{definition}

\begin{definition}[$0-1$ ratio]\label{def:0-1 ratio}
Let $x^n=(x_1,x_2,\cdots,x_n)$. Let $\gamma_k=\min\{\sum_{x^n\in f^{-1}(0)}\I_{\{x_k=0\}}, \sum_{x^n\in f^{-1}(0)}\I_{\{x_k=1\}}\}$, $k\in\{1,\dots,n\}$. The $0-1$ ratio at $k$-th position is $\gamma_k/(M-\gamma_k)$. Clearly, $0\le \gamma_k\le \lfloor\frac{M}{2} \rfloor$.
\end{definition}

\begin{definition}[ratio spectrum]\label{def:ratio spectrum}
The ratio spectrum of $f^{-1}(0)$ is defined by an integer sequence $\R_f=\{r_0,r_1,\cdots,r_{\lfloor\frac{M}{2}\rfloor}\}$, where $r_i=\sum_{k=1}^n\I_{\{\gamma_k=i\}}, i\in\{0,1,\cdots,\lfloor\frac{M}{2}\rfloor\}$. Clearly, $\sum_{i=0}^{\lfloor\frac{M}{2}\rfloor}r_i=n$.
\end{definition}

\begin{definition}[lexicographic ordering of ratio spectra]
The ratio spectrum $\R_{f}=\{r_0,r_1,\cdots,r_{\lfloor\frac{M}{2}\rfloor}\}$ is said to be (strictly)  greater than $\R_{f'}=\{r_0',r_1',\cdots,r'_{\lfloor\frac{M}{2}\rfloor}\}$, denoted by $\R_{f}\succ\R_{f'}$, if and only if $r_{j}>r_{j}'$ for some $j$ and $r_i=r_i'$ for all $i<j$. 
\end{definition}

First, we present the general formula of $F_f''(1/2)$ which is uniquely determined by ratio spectrum $\R_f$.
\begin{theorem}\label{theorem: second_deriv}
Given $n$, for any Boolean function $f$, the ratio spectrum $\R_f=\{r_0,r_1,\cdots,r_{\lfloor\frac{M}{2} \rfloor}\}$ uniquely determines $F_f''(1/2)$, that is,
\begin{align}
F_f''\left(\frac12\right)=&-4+\frac{4}{(2^n-M)M}\left(nM^2-4\sum_{t=0}^{\lfloor\frac{M}{2} \rfloor}(M-t)tr_t\right).\label{eq:second_deriv_formula}
\end{align}
\end{theorem}

\begin{IEEEproof}
See Appendix \ref{appendix: second_deriv derivation} for complete derivations.
\end{IEEEproof}

Theorem \ref{theorem: second_deriv} indicates that $F_f''(1/2)$ only depends on dimension $n$, cardinality $M$ of set $f^{-1}(0)$, and ratio spectrum $\R_f$. Here, we discover that ratio spectrum $\R_f$ acts as the ``structure'' of Boolean function $f$. 
\begin{corollary}\label{corollary: bigger ratio spectra}
For a given $n$, let $|f_1^{-1}(0)|=|f_2^{-1}(0)|=M, 1\le M\le 2^{n-1}$, for Boolean functions $f_1$ and $f_2$. If $\R_{f_1}\succ\R_{f_2}$, then $F_{f_1}''(1/2)>F_{f_2}''(1/2)$.
\end{corollary}

\begin{IEEEproof}
Let $\R_{f_1}=\{r_0, \cdots, r_i, \cdots, r_j, \cdots, r_{\lfloor\frac{M}{2} \rfloor}\}$ and $\R_{f_2}=\{r_0, \cdots, r_i-1, \cdots, r_j+1, \cdots, r_{\lfloor\frac{M}{2} \rfloor}\}$, where $0\le i<j\le\lfloor\frac{M}{2} \rfloor, r_i>0, r_j>0$. Obviously, $\R_{f_1}\succ \R_{f_2}$. By \eqref{eq:second_deriv_formula}, we have
\begin{align}
F_{f_1}''(1/2)-F_{f_2}''(1/2)=&\frac{16(j-i)(M-i-j)}{(S-M)M}>0.
\end{align}
In general, any other spectrum inequality can be established by successively constructing the above two ``adjacent'' ratio spectra.
\end{IEEEproof}

Corollary \ref{corollary: bigger ratio spectra} indicates that $F_f''(1/2)$ is an increasing function in ratio spectrum $\R_f$. Therefore, among Boolean functions with the same $|f^{-1}(0)|$, the one with the largest ratio spectrum maximizes $F_f''(1/2)$. The following theorem shows that the lex function is one type of functions with the largest ratio spectrum.

\begin{theorem}
For a given $n$, among all Boolean functions with the same $|f^{-1}(0)|$, the lex function is a function with the largest ratio spectrum $\R_f^*$.
\end{theorem}

\begin{IEEEproof}
It is enough to prove that when $f$ is lex, $\R_f\succeq \R_{f'}$ always holds, where $f'$ is any other Boolean function with the same $|f^{-1}(0)|$. This can be shown by contradiction.

Assume Boolean function $f$ is lex and $f'$ satisfies $\R_{f'}\succ \R_f$. This is only possible by first deleting bit $1$'s in $f^{-1}(0)$ and then performing any of the following operations:
\begin{itemize}
\item permutation: to permutate $x_k$'s among all $x^n\in f^{-1}(0)$;
\item flipping: to flip $x_k$ for each $x^n\in f^{-1}(0)$;
\item switching: to switch $x_i$ and $x_j$ for each $x^n\in f^{-1}(0)$.
  \end{itemize}
However, only deletion of bit $1$'s can result in a $\R_{f'}$ greater than $\R_f$, which will also result in repetitive binary sequences in ${f'}^{-1}(0)$. Since the latter consequence contradicts the definition of Boolean functions, this means that the lex function has the largest ratio spectrum. 
\end{IEEEproof}
Note that, when $|f^{-1}(0)|<2^{n-1}$ is fixed, there exist other Boolean functions that have the same largest ratio spectrum as the lex function. Therefore, the lex function can be referred to as \emph{a locally most informative function}. However, when $|f^{-1}(0)|=2^{n-1}$, the lex function reduces to dictator function $f(X^n)=X_1$. Since permutation and flipping still result in $f(X^n)=X_1$, and switching produces other dictator function $f(X^n)=X_i, 1<i\le n$, this implies that the dictator function is the only type of functions with the largest ratio spectrum. Next, we will show that the dictator function is also the only type of functions that can achieve $F_f''(1/2)=0$. Functions other than dictator functions can only result in $F''_f(1/2)<0$. Therefore, dictator functions are referred to as \emph{the globally most informative function}.

We are now ready to accomplish the last step of proving Lemma \ref{lemma: second_deriv}, the nonpositivity of $F_f''(1/2)$. To achieve this, we define
\begin{align}
W_f(M)\triangleq&\sum_{t=0}^{\lfloor\frac{M}{2}\rfloor}(M-t)tr_t \label{eq: constant W} \\
g(n)\triangleq&-4+\frac{4\Big(nM^2-4W_f(M)\Big)}{(2^n-M)M}.\label{eq: g(n)}
\end{align}
Note that $g(n)$ is exactly $F_f''(1/2)$ in \eqref{eq:second_deriv_formula}. Our goal is to show that if Boolean function $f$ is lex, (i) $W_f(M)$ remains constant in $n$ if $n\ge\lceil\log_2M \rceil$; (ii) $g(n)$ is a monotonically decreasing function for $n\ge \log_2M+1$. 

\begin{proposition}\label{proposition: constant}
With Boolean function $f$ being lex, $W_f(M)$ in \eqref{eq: constant W} is constant in $n$ if $n\ge\lceil\log_2M \rceil$.
\end{proposition}

\begin{IEEEproof}
It is equivalent to examining the case when $0\le M\le 2^n$. Since $f$ is lex, it can be verified that
\begin{align*}
\sum_{x^n\in f^{-1}(0)}\I_{\{x_i=1\}}=\sum_{k=0}^{2^{n-i}-1}\left\lfloor\frac{M+k}{2^{n+1-i}}\right\rfloor.
\end{align*}
for $i\in\{1,2,\cdots,n\}$ and $x^n=(x_1,x_2,\cdots,x_n)$. Obviously, $\sum_{x^n\in f^{-1}(0)}\I_{\{x_i=1\}}=0$ if $i\le n-\lceil\log_2M\rceil$.

Therefore, according to Definition \ref{def:0-1 ratio} and Definition \ref{def:ratio spectrum},
\begin{align}
W_f(M)&=\sum_{i=1}^n\sum_{t=0}^{\lfloor\frac{M}{2}\rfloor}(M-t)t\cdot\I_{\{\gamma_i=t\}}\notag\\
  &=\sum_{i=1}^n\left(M-\sum_{x^n\in f^{-1}(0)}\I_{\{x_i=1\}}\right)\left(\sum_{x^n\in f^{-1}(0)}\I_{\{x_i=1\}}\right)\notag\\
  &=\sum_{i=1}^n\left(M-\sum_{k=0}^{2^{n-i}-1}\left\lfloor\frac{M+k}{2^{n+1-i}}\right\rfloor\right)\left(\sum_{k=0}^{2^{n-i}-1}\left\lfloor\frac{M+k}{2^{n+1-i}}\right\rfloor\right),\label{eq: sum of weights}
\end{align}
which will be constant as long as $n\ge\lceil\log_2M \rceil$.
\end{IEEEproof}

\begin{proposition}\label{proposition: nonpositivity}
Assume $f$ is lex and $n\ge \log_2M+1$ (since it suffices to examine $1\le M\le 2^{n-1}$ by Theorem \ref{theorem:complementary function}). We have
\begin{align}
M^2\log_2M\le 4W_f(M)<M^2\left(\log_2M+\frac{2\ln2-1}{2\ln2}\right) \label{eq: constant part}
\end{align}
and
\begin{align}
g(n)\le g(\log_2M+1)\le 0,\label{eq:g function}
\end{align}
where in \eqref{eq:g function}, the first equality holds if and only if $n=\log_2M+1$ and the second equality holds if and only if $M$ is a power of $2$. Consequently, $F_f''(1/2)=0$ if and only if $f$ is the dictator function.
\end{proposition}

\begin{IEEEproof}
We first prove \eqref{eq: constant part}. By replacing $i$ with $n+1-i$ in \eqref{eq: sum of weights}, we have
\begin{align}
W_f(M)=&\sum_{i=1}^n\left(M-\sum_{k=0}^{2^{i-1}-1}\left\lfloor\frac{M+k}{2^{i}}\right\rfloor\right)\left(\sum_{k=0}^{2^{i-1}-1}\left\lfloor\frac{M+k}{2^{i}}\right\rfloor\right).\label{eq:change variable}
\end{align}
It can be verified that for $i=1,2,\cdots,n$,
\begin{align}\label{eq:mod cases}
\sum_{k=0}^{2^{i-1}-1}\left\lfloor\frac{M+k}{2^{i}}\right\rfloor=\begin{cases}
2^{i-1}\left\lfloor\frac12+\frac{M}{2^i} \right\rfloor,&\text{if }\left\lfloor\frac{M}{2^{i-1}}\right\rfloor\equiv0\mod 2\\
M-2^{i-1}\left\lfloor\frac12+\frac{M}{2^i} \right\rfloor,&\text{if }\left\lfloor\frac{M}{2^{i-1}}\right\rfloor\equiv1\mod 2.
\end{cases}
\end{align}
Therefore, substituting \eqref{eq:mod cases} into \eqref{eq:change variable} yields
\begin{align*}
W_f(M)=&\sum_{i=1}^{n}\left(M-2^{i-1}\left\lfloor\frac12+\frac{M}{2^i} \right\rfloor\right)\left(2^{i-1}\left\lfloor\frac12+\frac{M}{2^i} \right\rfloor\right)\\
\triangleq&a(M-1).
\end{align*}
By \cite{OEIS_Sequence}, $a(M)$ can be computed recursively as follows.
\begin{align*}
a(0)&=0\\
a(2M)&=2a(M)+2a(M-1)+M(M+1)\\
a(2M+1)&=4a(M)+(M+1)^2,
\end{align*}
for $M\in\Z$. Thus by induction, it can be shown that
\begin{align}
M^2\log_2M\le 4a(M-1)<M^2\left(\log_2M+\frac{2\ln2-1}{2\ln2}\right), \label{eq: induction inequality}
\end{align}
where the equality holds if and only if $M$ is a power of $2$. Complete induction steps can be found in Appendix \ref{appendix: bounds for W(M)}.

Now we prove \eqref{eq:g function}. Since $f$ is lex and $n\ge\log_2M+1$, making $n$ a continuous variable, $g'(n)$ is given by
\begin{align}
g'(n)=&4\cdot\frac{(2^n-M)M^2-(nM^2-4W_f(M))2^n\ln 2}{(2^n-M)^2M}.
\end{align}
In order to show that $g'(n)<0$ for $n\ge \log_2 M+1$, we need to show
\begin{align}
4W_f(M)<&\min_{n\ge\log_2M+1}\left\{\frac{M^3}{2^n\ln 2}+\left(n-\frac1{\ln2}\right)M^2\right\}\notag\\
=&M^2\left(\log_2M+\frac{2\ln2-1}{2\ln2}\right),
\end{align}
which has just been corroborated in \eqref{eq: induction inequality}. Therefore, $g(n)$ is a monotonically decreasing function and by \eqref{eq: induction inequality}
\begin{align}
g(n)\le&g(\log_2M+1)\notag\\
  =&-4+\frac{4\Big((\log_2M+1)M^2-4W_f(M)\Big)}{(2^{\log_2M+1}-M)M}\notag\\
  \le& -4+\frac{4}{M^2}\Big[(\log_2M+1)M^2-M^2\log_2M\Big]\notag\\
  =&0.
\end{align}
The proof is completed.
\end{IEEEproof}
Proposition \ref{proposition: nonpositivity} implies that for any Boolean function $f$ being lex, $F_f''(1/2)\le 0$, where the equality holds if and only if $M$ is a power of $2$ and $n=\log_2 M+1$, suggesting that $f$ can only be the dictator function. By the aforementioned most informative Boolean function argument, we conclude that dictator function is the globally most informative function in the high noise regime, among all possible choices of $f$.

\section{Discussion}\label{sec: discussion}
In this paper, we establish Conjecture \ref{conjecture: mutual info} in the high noise regime by looking at the derivatives of  $F_f(\alpha)$. The limitations and future directions of this approach are as follows:

A limitation is that  $\delta_n$ in Theorem \ref{theorem: main result} is dimensionally dependent on $n$, which weakens our result compared to Samorodnitsky's, where $\delta$ is a universal, dimension-free constant.

As a future direction, we note that to the best of our knowledge, current numerical exhaustive search indicates that there are only two possible shapes of $F_f(\alpha), \alpha\in(0,1)$, as depicted in Fig. \ref{fig:example of F and T}.  This seems promising for future investigation. The calculus-based approach presented in this paper could possibly facilitate a new approach to tackle Conjecture \ref{conjecture: mutual info}. Alternatively, the new function $F_f(\alpha)$ for $\alpha\in(0,1)$ might be analyzed using other techniques.

\section*{Acknowledgment}

We are grateful to Jiange Li for valuable comments on an earlier version of this paper.

\appendices

\section{Proof of Theorem \ref{theorem: second_deriv}}
\label{appendix: second_deriv derivation}

Assume the natural logarithm. According to \eqref{eq: definition of F_f},
\begin{align}
F_f(\alpha)=&H(\alpha)-\frac1S\sum_{y^n\in\setS}H\Big(\Prob\{f(X^n)=0|y^n,\alpha\}\Big)\\
  =&H(\alpha)-\frac1S\sum_{y^n\in\setS}H\left(\sum_{x^n\in f^{-1}(0)}\alpha^{d(x^n,y^n)}\lalpha^{n-d(x^n,y^n)}\right).
\end{align}
Therefore, the first derivative of $F_f(\alpha)$ is given as follows.
{\begin{align}
F_f'(\alpha)=&\log\frac{\lalpha}{\alpha}+\frac1S\sum_{y^n\in\setS}\frac{\partial \Prob\{f(X^n)=0|y^n,\alpha\}}{\partial \alpha}\log\frac{\Prob\{f(X^n)=0|y^n,\alpha\}}{1-\Prob\{f(X^n)=0|y^n,\alpha\}}\\
F_f''(\alpha)=&\frac{-1}{\alpha\lalpha}+\frac1S\sum_{y^n\in\setS}\Bigg\{\frac{\partial^2 \Prob\{f(X^n)=0|y^n,\alpha\}}{\partial \alpha^2}\log\frac{\Prob\{f(X^n)=0|y^n,\alpha\}}{1-\Prob\{f(X^n)=0|y^n,\alpha\}}\notag\\
\phantom{}&+\frac{\left(\frac{\partial \Prob\{f(X^n)=0|y^n,\alpha\}}{\partial \alpha}\right)^2}{(1-\Prob\{f(X^n)=0|y^n,\alpha\})\Prob\{f(X^n)=0|y^n,\alpha\}}\Bigg\},
\end{align}}
where
\begin{align}
\Prob\{f(X^n)=0|y^n,\alpha\}=&\sum_{x^n\in f^{-1}(0)}\alpha^{d(x^n,y^n)}\lalpha^{n-d(x^n,y^n)}\\
\frac{\partial \Prob\{f(X^n)=0|y^n,\alpha\}}{\partial \alpha}=&\sum_{x^n\in f^{-1}(0)}(d(x^n,y^n)-n\alpha)\alpha^{d(x^n,y^n)-1}\lalpha^{n-1-d(x^n,y^n)}\label{eq: first_deriv_of_prob}\\
\frac{\partial^2 \Prob\{f(X^n)=0|y^n,\alpha\}}{\partial\alpha^2}=&\sum_{x^n\in f^{-1}(0)}\Big(d(x^n,y^n)(d(x^n,y^n)-1)+2(1-n)d(x^n,y^n)\alpha+(n^2-n)\alpha^2\Big)\notag\\
\phantom{=}&\cdot\alpha^{d(x^n,y^n)-2}\lalpha^{n-2-d(x^n,y^n)}.\label{eq: second_deriv_of_prob}
\end{align}

For convenience, let $|f^{-1}(0)|=M$. When $\alpha=1/2$, for all $y^n\in\setS$, $\Prob\{f(X^n)=0|y^n,\alpha=1/2\}=M/S$. Therefore, using Lemma \ref{lemma: conditional identity} and \eqref{eq: first_deriv_of_prob}, 
\begin{align}
F_f''\left(\frac12\right)=&-4+\frac{S}{(S-M)M}\sum_{y^n\in\setS}\left(\frac{\partial \Prob\{f(X^n)=0|y^n,\alpha\}}{\partial \alpha}\right)^2\Bigg|_{\alpha=1/2}\\
  =&-4+\frac{4}{(S-M)MS}\sum_{y^n\in\setS}\left(\sum_{x^n\in f^{-1}(0)}\Big(2d(x^n,y^n)-n\Big)\right)^2\\
  =&-4+\frac{4}{(S-M)MS}\sum_{y^n\in\setS}\left(4\Big(\sum_{x^n\in f^{-1}(0)}d(x^n,y^n)\Big)^2-4nM\sum_{x^n\in f^{-1}(0)}d(x^n,y^n)+n^2M^2\right)\\
  =&-4+\frac{4}{(S-M)MS}\left(4\sum_{y^n\in\setS}\Big(\sum_{x^n\in f^{-1}(0)}d(x^n,y^n)\Big)^2-4nM\cdot\frac12nMS+n^2M^2S\right)\\
  =&-4+\frac{4}{(S-M)MS}\left(4\sum_{y^n\in\setS}\Big(\sum_{x^n\in f^{-1}(0)}d(x^n,y^n)\Big)^2-n^2M^2S\right). \label{eq: second_deriv_v1}
\end{align}

At this point, we define the following notation which will simplify the above derivation.
\begin{align}
\setC_t\triangleq&\left\{i: \sum_{x^n\in f^{-1}(0)}\I_{\{x_i=1\}}=t,\ \forall i\in\{1,\cdots,n\}\right\},\quad t=0,1,\dots,M\\
C_t\triangleq&|\setC_t|\\
a_{t,y^n}\triangleq&\sum_{i\in\setC_t}\I_{\{y_i=1\}}\\
a_{y^n}\triangleq&\sum_{i=1}^n\I_{\{y_i=1\}}=\sum_{t=0}^M\sum_{i\in\setC_t}\I_{\{y_i=1\}}=\sum_{t=0}^Ma_{t,y^n}.
\end{align}
Essentially, the above notation considers the \emph{weight spectrum} $\{C_t\}_{t=0}^M$ of $f^{-1}(0)$. Therefore, from \eqref{eq: second_deriv_v1}
\begin{align}
\sum_{x^n\in f^{-1}(0)}d(x^n,y^n)=&\sum_{t=0}^M\Big((M-t)a_{t,y^n}+t(C_t-a_{t,y^n})\Big)\\
  =&Ma_{y^n}+\sum_{t=0}^Mt(C_t-2a_{t,y^n})\label{eq: sum_of_squares}
\end{align}
and 
\begin{align}
\phantom{=}&\sum_{y^n\in\setS}\Big(\sum_{x^n\in f^{-1}(0)}d(x^n,y^n)\Big)^2\notag\\
=&\sum_{y^n\in\setS}\left(Ma_{y^n}+\sum_{t=0}^Mt(C_t-2a_{t,y^n})\right)^2\\
  =&\sum_{y^n\in\setS}\left(M^2a^2_{y^n}+\left(\sum_{t=0}^Mt(C_t-2a_{t,y^n})\right)^2+2Ma_{y^n}\sum_{t=0}^Mt(C_t-2a_{t,y^n})\right)\\
  =&M^2\sum_{y^n\in\setS}a^2_{y^n}+\sum_{y^n\in\setS}\left(\left(\sum_{t=0}^MtC_t\right)^2+4\left(\sum_{t=0}^Mta_{t,y^n}\right)^2-4\left(\sum_{t=0}^MtC_t\right)\left(\sum_{t=0}^Mta_{t,y^n}\right) \right)\notag\\
  \phantom{}&+2M\sum_{t=0}^M\sum_{y^n\in\setS}\Big(tC_ta_{y^n}-2ta_{y^n}a_{t,y^n}\Big) \label{eq: complex terms}\\
  =&\frac14n(n+1)M^2S+S\sum_{t=0}^Mt^2C_t-MS\sum_{t=0}^MtC_t\label{eq: simple terms}\\
  =&\frac14n(n+1)M^2S-S\sum_{t=0}^M(M-t)tC_t,\label{eq: simplified_results}
\end{align}
where \eqref{eq: complex terms} to \eqref{eq: simple terms} follows from Lemma \ref{lemma: properties of a's}. Thus,
substituting \eqref{eq: simplified_results} into \eqref{eq: second_deriv_v1} gives the desired expression
\begin{align}
F_f''\left(\frac12\right)=&-4+\frac{4}{(S-M)MS}\left(n(n+1)M^2S-4S\sum_{t=0}^M(M-t)tC_t-n^2M^2S\right)\\
  =&-4+\frac{4}{(S-M)M}\left(nM^2-4\sum_{t=0}^M(M-t)tC_t\right)\label{eq: second_deriv_C_ts}\\
  =&-4+\frac{4}{(2^n-M)M}\left(nM^2-4\sum_{t=0}^{\lfloor\frac{M}{2}\rfloor}(M-t)tr_t\right), \label{eq: second_deriv_R_ts}
\end{align}
where \eqref{eq: second_deriv_C_ts} to \eqref{eq: second_deriv_R_ts} is from that $r_t=C_t+C_{M-t}$ and that $(M-t)t$ remains the same for $C_t$ and $C_{M-t}$.

\begin{lemma}\label{lemma: conditional identity}
If $Y^n\in\setS$ is equiprobable, for any $\alpha\in(0,1)$, we have
\begin{align}
\sum_{y^n\in\setS}\Prob\{f(X^n)=0|y^n,\alpha\}=&|f^{-1}(0)|\\
\sum_{y^n\in\setS}\frac{\partial^i \Prob\{f(X^n)=0|y^n,\alpha\}}{\partial\alpha^i}=&0,\ (i\ge1). \label{eq: sum_to_zero}
\end{align}
\end{lemma}
\begin{IEEEproof}
Since $Y^n\in\setS$ is equiprobable, we have
\begin{align}
\sum_{y^n\in\setS}\Prob\{f(X^n)=0|y^n,\alpha\}=&\sum_{y^n\in\setS}\Prob\{f(X^n)=0|Y^n=y^n\}\Prob\{Y^n=y^n\}\cdot\frac{1}{\Prob\{Y^n=y^n\}}\\
  =&S\sum_{y^n\in\setS}\Prob\{f(X^n)=0|Y^n=y^n\}\Prob\{Y^n=y^n\}\\
  =&S\Prob\{f(X^n)=0\}\\
  =&|f^{-1}(0)|,
\end{align}
which immediately implies \eqref{eq: sum_to_zero}.

\end{IEEEproof}

\begin{lemma}\label{lemma: properties of a's}
With the notation defined above, 
\begin{align}
\sum_{y^n\in\setS}a_{y^n}^2=&\sum_{k=0}^n\binom{n}{k}k^2=n(n+1)\cdot2^{n-2}\label{eq: A1}\\
\sum_{y^n\in\setS}a_{t,y^n}^2=&\sum_{k=0}^n\sum_{k_1=0}^k\binom{n-C_t}{k-k_1}\binom{C_t}{k_1}k_1^2=C_t(C_t+1)\cdot2^{n-2}\label{eq: A2}\\
\sum_{y^n\in\setS}a_{y^n}a_{t,y^n}=&\sum_{k=0}^nk\binom{n-1}{k-1}C_t=C_t(n+1)\cdot2^{n-2}\label{eq: A3}\\
\sum_{y^n\in\setS}a_{t_1,y^n}a_{t_2,y^n}=&\sum_{k=0}^n\sum_{k_1=0}^k\sum_{k_2=0}^{k-k_1}\binom{n-C_{t_1}-C_{t_2}}{k-k_1-k_2}\binom{C_{t_1}}{k_1}\binom{C_{t_2}}{k_2}k_1k_2=C_{t_1}C_{t_2}\cdot2^{n-2},\ (t_1\ne t_2).\label{eq: A4}
\end{align}
\end{lemma}
\begin{IEEEproof}
The leftmost terms above are combinatorial problems which can be solved by enumerating the weights accordingly and then calculating the summation. Thus, \eqref{eq: A1} is established by Lemma \ref{lemma: expasion of (1+x)^n}. \eqref{eq: A2} is established by Lemma \ref{lemma: expasion of (1+y)^m(x+y)^n}. \eqref{eq: A3} comes from $a_{t,y^n}=\binom{n-1}{k-1}C_t$ when $a_{y^n}=k$ is fixed and then follows from Lemma \ref{lemma: expasion of (1+x)^n}. \eqref{eq: A4} is established by Lemma \ref{lemma: expansion of (1+x)^m(y+x)^n(z+x)^t}.
\end{IEEEproof}

\begin{lemma}\label{lemma: expasion of (1+x)^n}
The expansion of $(1+x)^n$, $n\in\mathbb{Z}^+$, yields the following identities
\begin{align}
\sum_{k=0}^n\binom{n}{k}=&2^n\\
\sum_{k=0}^n\binom{n}{k}k=&n\cdot2^{n-1}\\
\sum_{k=0}^n\binom{n}{k}k^2=&n(n+1)\cdot2^{n-2}.
\end{align}
\end{lemma}
\begin{IEEEproof}
All identities above can be derived from
\begin{align}
(1+x)^n=\sum_{k=0}^n\binom{n}{k}x^k
\end{align}
by taking derivatives with respect to $x$ and evaluating at $x=1$.
\end{IEEEproof}

\begin{lemma}\label{lemma: expasion of (1+y)^m(x+y)^n}
The expansion of $(1+y)^m(x+y)^n$, $m,n\in\mathbb{Z}^+$, yields the following identities
\begin{align}
\sum_{r=0}^{m+n}\sum_{k=0}^r\binom{m}{k}\binom{n}{r-k}k=&m\cdot2^{m+n-1}\\
\sum_{r=0}^{m+n}\sum_{k=0}^r\binom{m}{k}\binom{n}{r-k}k^2=&m(m+1)\cdot2^{m+n-2}.
\end{align}
\end{lemma}
\begin{IEEEproof}
The above identities can be derived from
\begin{align}
(1+y)^m(x+y)^n=&\left(\sum_{i=0}^m\binom{m}{i}y^i\right)\left(\sum_{j=0}^n\binom{n}{j}x^{n-j}y^j\right)\\
  =&\sum_{r=0}^{m+n}\left(\sum_{k=0}^r\binom{m}{k}\binom{n}{r-k}x^{n-r+k}\right)y^r
\end{align}
by taking derivatives with respect to $x$, evaluating at $x=1$, and evaluating at $y=1$ accordingly. 
\end{IEEEproof}

\begin{lemma}\label{lemma: expansion of (1+x)^m(y+x)^n(z+x)^t}
The expansion of $(1+x)^m(y+x)^n(z+x)^t$, $m,n,t\in\mathbb{Z}^+$, yields the following identity
\begin{align}
\sum_{r=0}^{m+n+t}\sum_{k=0}^r\sum_{l=0}^{r-k}\binom{m}{r-k-l}\binom{n}{k}\binom{t}{l}kl=nt\cdot2^{m+n+t-2}.
\end{align}
\end{lemma}
\begin{IEEEproof}
The above identity can be derived from
\begin{align}
(1+x)^m(y+x)^n(z+x)^t=&\left(\sum_{i=0}^m\binom{m}{i}x^i\right)\left(\sum_{j=0}^n\binom{n}{j}y^{n-j}x^j\right)\left(\sum_{k=0}^t\binom{t}{k}z^{t-k}x^k\right)\\
    =&\sum_{r=0}^{m+n+t}\left(\sum_{k=0}^{r}\sum_{l=0}^{r-k}\binom{m}{r-k-l}\binom{n}{k}\binom{t}{l}y^{n-k}z^{t-l}\right)x^r
\end{align}
by taking the derivatives with respect to $y$, taking the derivatives with respect to $z$, evaluating at $x=1$, evaluating at $y=1$, evaluating at $z=1$ accordingly.
\end{IEEEproof}

\section{Proof of Inequality \eqref{eq: induction inequality}}
\label{appendix: bounds for W(M)}

Define for $m\ge0$,
\begin{align}
a(m)\triangleq \sum_{i=1}^{\lceil\log_2(m+1) \rceil}\left(m+1-2^{i-1}\left\lfloor\frac12+\frac{m+1}{2^i} \right\rfloor\right)\left(2^{i-1}\left\lfloor\frac12+\frac{m+1}{2^i} \right\rfloor\right).
\end{align}

By \cite{OEIS_Sequence}, $a(m)$ can be computed recursively as follows.
\begin{align}
a(0)&=0\\
a(2m)&=2a(m)+2a(m-1)+m(m+1)\label{eq:even}\\
a(2m+1)&=4a(m)+(m+1)^2.\label{eq:odd}
\end{align}
Thus it is equivalent to proving that, for any $m\in\mathbb{Z}^+$,
\begin{align}
\frac14(m+1)^2\log_2(m+1)\le a(m)<\frac14(m+1)^2\left(\log_2(m+1)+\frac{2\ln2-1}{2\ln2}\right).\label{eq: inequality of sequence}
\end{align}

When $m=0,1$, it can be verified that \eqref{eq: inequality of sequence} holds. Assume \eqref{eq: inequality of sequence} holds for any $m\le M$, where $M$ is a constant and $M\ge1$. Consider $m=M+1$. We first establish the lower bound.

1) If $M$ is odd, then $m=M+1$ is even and we apply \eqref{eq:even}.
\begin{align}
\phantom{=}&a(M+1)\notag\\
=&2a\left(\frac{M+1}{2}\right)+2a\left(\frac{M-1}{2}\right)+\left(\frac{M+1}{2}\right)\left(\frac{M+3}{2}\right)\\
\ge&\frac12\left(\frac{M+3}{2}\right)^2\log_2\left(\frac{M+3}{2}\right)+\frac12\left(\frac{M+1}{2}\right)^2\log_2\left(\frac{M+1}{2}\right)+\left(\frac{M+1}{2}\right)\left(\frac{M+3}{2}\right)\label{eq:285}\\
\ge&\frac14(M+2)^2\log_2(M+2)\label{eq:286},
\end{align}
where \eqref{eq:285} to \eqref{eq:286} follows from the following fact. 

Define
\begin{align}
f(x)\triangleq&\frac12\left(\frac{x+3}{2}\right)^2\log_2\left(\frac{x+3}{2}\right)+\frac12\left(\frac{x+1}{2}\right)^2\log_2\left(\frac{x+1}{2}\right)+\left(\frac{x+1}{2}\right)\left(\frac{x+3}{2}\right)\notag\\
\phantom{=}&-\frac14(x+2)^2\log_2(x+2)\\
    =&\frac18 (x+3)^2\log_2(x+3)+\frac18(x+1)^2\log_2(x+1)-\frac14 (x+2)^2\log_2(x+2)-\frac12.
\end{align}
Thus,
\begin{align}
f'(x)=\frac14 (x+3)\log_2(x+3)+\frac14(x+1)\log_2(x+1)-\frac12(x+2)\log_2(x+2)\ge0
\end{align}
which follows from that $x\log_2x$ is convex and Jensen's inequality. Hence,
\begin{align}
f(x)\ge f(0)=\frac38\log_2\left(\frac{27}{16}\right)>0.
\end{align}

2) If $M$ is even, then $m=M+1$ is odd and we apply \eqref{eq:odd},
\begin{align}
a(M+1)&=4a\left(\frac M2\right)+\left(\frac{M+2}{2}\right)^2\\
    &\ge \left(\frac{M+2}{2}\right)^2\log_2\left(\frac{M+2}{2}\right)+\left(\frac{M+2}{2}\right)^2\\
    &=\frac14\left(M+2\right)^2\log_2\left(M+2\right). \label{eq: 95}
\end{align}
Combining \eqref{eq:286} and \eqref{eq: 95}, the lower bound in \eqref{eq: inequality of sequence} holds for $m=M+1$.

Next, we establish the upper bound by proving a tighter upper bound for $m\ge6$, given as follows (For $1\le m\le 5$, we can verify that the original upper bound in \eqref{eq: inequality of sequence} holds).
\begin{align}
a(m)\le\frac14 (m+1)^2\left(\log_2(m+1)+\frac{(m-1)b}{m}\right).\label{eq:tight_bound}
\end{align}
where $b\triangleq\frac{2\ln2-1}{2\ln2}$. Clearly, the upper bound in \eqref{eq:tight_bound} implies the upper bound in \eqref{eq: inequality of sequence}.

First, it can be verified that when $m=6$ and $m=7$, \eqref{eq:tight_bound} holds. Assume \eqref{eq:tight_bound} holds for $m\le M$, where $M$ is a constant and $M\ge7$. Consider $m=M+1$.

1) If $M$ is odd, then $m=M+1$ is even and we apply \eqref{eq:even}.
\begin{align}
\phantom{=}&a(M+1)\notag\\
=&2a\left(\frac{M+1}{2}\right)+2a\left(\frac{M-1}{2}\right)+\left(\frac{M+1}{2}\right)\left(\frac{M+3}{2}\right)\\
\le&\frac12\left(\frac{M+3}{2}\right)^2\left(\log_2\left(\frac{M+3}{2}\right)+\frac{(M-1)b}{M+1}\right)+\frac12\left(\frac{M+1}{2}\right)^2\left(\log_2\left(\frac{M+1}{2}\right)+\frac{(M-3)b}{M-1}\right)\notag\\
\phantom{+}&+\left(\frac{M+1}{2}\right)\left(\frac{M+3}{2}\right)\label{eq:273}\\
<&\frac14(M+2)^2\left(\log_2(M+2)+\frac{Mb}{M+1}\right),\label{eq:274}
\end{align}
where \eqref{eq:273} to \eqref{eq:274} follows from the following fact. 

Define
\begin{align}
f(x)\triangleq& \frac12\left(\frac{x+3}{2}\right)^2\left(\log_2\left(\frac{x+3}{2}\right)+\frac{(x-1)b}{x+1}\right)+\frac12\left(\frac{x+1}{2}\right)^2\left(\log_2\left(\frac{x+1}{2}\right)+\frac{(x-3)b}{x-1}\right)\notag\\
\phantom{+}&+\left(\frac{x+1}{2}\right)\left(\frac{x+3}{2}\right)-\frac14(x+2)^2\left(\log_2(x+2)+\frac{bx}{x+1}\right)\\
    =&\frac18(x+3)^2\log_2(x+3)+\frac18(x+1)^2\log_2(x+1)-\frac14(x+2)^2\log_2(x+2)+\frac{(x+1)(x+3)}{4}\notag\\
    \phantom{=}&-\frac{\big[(1-b)x+(1+b)\big](x+3)^2}{8(x+1)}-\frac{\big[(1-b)x+(3b-1)\big](x+1)^2}{8(x-1)}-\frac{bx(x+2)^2}{4(x+1)}.
\end{align}
Thus,
\begin{align}
f'(x)=&\frac14(x+3)\log_2(x+3)+\frac14(x+1)\log_2(x+1)-\frac12(x+2)\log_2(x+2)+\frac{x+2}{2}\notag\\
    &-\frac{\big[(1-b)x^2+(2-b)x+(1-2b)\big](x+3)}{4(x+1)^2}-\frac{\big[(1-b)x^2+(3b-2)x-4b\big](x+1)}{4(x-1)^2}\notag\\
    &-\frac{(2bx^2+3bx+2b)(x+2)}{4(x+1)^2}.
\end{align}
It can be verified that $f'(x)$ is a monotonically decreasing function of $x$ and $f'(x)<0$ as $x\ge7$. Thus, for any $m\in\Z$, we have
\begin{align}
f(m)\le f(7)\approx -0.00574<0.
\end{align}

2) If $M$ is even, then $m=M+1$ is odd and we apply \eqref{eq:odd},
\begin{align}
a(M+1)&=4a\left(\frac M2\right)+\left(\frac{M+2}{2}\right)^2\\
    &\le\left(\frac{M+2}{2}\right)^2\left(\log_2\left(\frac{M+2}{2}\right)+\frac{(M-2)b}{M}\right)+\left(\frac{M+2}{2}\right)^2\\
    &=\frac14\left(M+2\right)^2\left(\log_2\left(M+2\right)+\frac{(M-2)b}{M}\right)\\
    &<\frac14\left(M+2\right)^2\left(\log_2\left(M+2\right)+\frac{Mb}{M+1}\right). \label{eq: 108}
\end{align}
Therefore, \eqref{eq:tight_bound} holds for $m=M+1$. Combining \eqref{eq:274} and \eqref{eq: 108}, the tighter upper bound in \eqref{eq:tight_bound} holds for $m=M+1$, implying that the upper bound in \eqref{eq: inequality of sequence} holds.

In summary, both the lower bound and upper bound hold in \eqref{eq: inequality of sequence}.

\bibliography{references}

\begin{thebibliography}{10}
\providecommand{\url}[1]{#1}
\csname url@samestyle\endcsname
\providecommand{\newblock}{\relax}
\providecommand{\bibinfo}[2]{#2}
\providecommand{\BIBentrySTDinterwordspacing}{\spaceskip=0pt\relax}
\providecommand{\BIBentryALTinterwordstretchfactor}{4}
\providecommand{\BIBentryALTinterwordspacing}{\spaceskip=\fontdimen2\font plus
\BIBentryALTinterwordstretchfactor\fontdimen3\font minus
  \fontdimen4\font\relax}
\providecommand{\BIBforeignlanguage}[2]{{%
\expandafter\ifx\csname l@#1\endcsname\relax
\typeout{** WARNING: IEEEtran.bst: No hyphenation pattern has been}%
\typeout{** loaded for the language `#1'. Using the pattern for}%
\typeout{** the default language instead.}%
\else
\language=\csname l@#1\endcsname
\fi
#2}}
\providecommand{\BIBdecl}{\relax}
\BIBdecl

\bibitem{Courtade_2014}
T.~A. Courtade and G.~R. Kumar, ``Which boolean functions maximize mutual
  information on noisy inputs?'' \emph{IEEE Transactions on Information
  Theory}, vol.~60, no.~8, pp. 4515--4525, Aug 2014.

\bibitem{Wyner_1973}
A.~Wyner and J.~Ziv, ``A theorem on the entropy of certain binary sequences and
  applications--i,'' \emph{IEEE Transactions on Information Theory}, vol.~19,
  no.~6, pp. 769--772, November 1973.

\bibitem{Erkip_1998}
E.~Erkip and T.~M. Cover, ``The efficiency of investment information,''
  \emph{IEEE Transactions on Information Theory}, vol.~44, no.~3, pp.
  1026--1040, May 1998.

\bibitem{Ordentlich_2016}
O.~Ordentlich, O.~Shayevitz, and O.~Weinstein, ``An improved upper bound for
  the most informative boolean function conjecture,'' in \emph{2016 IEEE
  International Symposium on Information Theory (ISIT)}, July 2016, pp.
  500--504.

\bibitem{Samorodnitsky_2016}
A.~Samorodnitsky, ``On the entropy of a noisy function,'' \emph{IEEE
  Transactions on Information Theory}, vol.~62, no.~10, pp. 5446--5464, Oct
  2016.

\bibitem{Pichler_2016}
G.~Pichler, G.~Matz, and P.~Piantanida, ``A tight upper bound on the mutual
  information of two boolean functions,'' in \emph{2016 IEEE Information Theory
  Workshop (ITW)}, Sept 2016, pp. 16--20.

\bibitem{Kindler_2015}
\BIBentryALTinterwordspacing
G.~Kindler, R.~O'Donnell, and D.~Witmer, ``Remarks on the most informative
  function conjecture at fixed mean.'' [Online]. Available:
  \url{http://arxiv.org/abs/1506.03167}
\BIBentrySTDinterwordspacing

\bibitem{Anatharam_2013}
V.~Anantharam, A.~A. Gohari, S.~Kamath, and C.~Nair, ``On hypercontractivity
  and the mutual information between boolean functions,'' in \emph{2013 51st
  Annual Allerton Conference on Communication, Control, and Computing
  (Allerton)}, Oct 2013, pp. 13--19.

\bibitem{Huleihel_2017}
W.~Huleihel and O.~Ordentlich, ``How to quantize $n$ outputs of a binary
  symmetric channel to $n-1$ bits?'' in \emph{2017 IEEE International Symposium
  on Information Theory (ISIT)}, June 2017, pp. 91--95.

\bibitem{Li_2018}
J.~Li and M.~M\'edard, ``Boolean functions: Noise stability, non-interactive
  correlation, and mutual information,'' in \emph{2018 IEEE International
  Symposium on Information Theory (ISIT)}, June 2018, pp. 266--270.

\bibitem{OEIS_Sequence}
A.~Kundgen, ``The on-line encyclopedia of integer sequences, 2003, sequence
  a022560.''

\end{thebibliography}
\end{document}